\documentclass[twocolumn,showpacs,%
  nofootinbib,aps,superscriptaddress,%
  eqsecnum,prd,notitlepage,showkeys,10pt]{revtex4-1}

\usepackage{amssymb}
\usepackage{amsmath}
\usepackage{graphicx}
\usepackage{dcolumn}
\usepackage{hyperref}
\usepackage{mathtools}
\usepackage[utf8x]{inputenc}

\begin{document}

\title{Securing the Control-plane Channel and Cache of Pull-based ID/LOC Protocols}
\author{Paul Almasan*, Jordi Paillissé*, Alberto Rodriguez-Natal\textsuperscript{\S}, Pere Barlet-Ros*, Florin Coras\textsuperscript{\S} \\ Vina Ermagan\textsuperscript{\S}, Fabio Maino\textsuperscript{\S}, Albert Cabellos-Aparicio*\\
* Universitat Politècnica de Catalunya\\
Barcelona,Spain\\
\textsuperscript{\S} Cisco Systems\\
San Jose, CA, USA\\}  
\noaffiliation

\begin{abstract}
Pull-based ID/LOC split protocols, such as LISP (RFC6830), retrieve mappings from a mapping system to encapsulate and forward packets. This is done by means of a control-plane channel. In this short paper we describe three attacks against this channel (Denial-of-Service and overflowing) as well as the against the local cache used to store such mappings. We also provide a solution against such attacks that implements a per-source rate-limiter using a Count-Min Sketch data-structure.
\end{abstract}

\maketitle

\section{Introduction}

ID/LOC split protocols build on top of the basic idea of creating two separate namespaces: overlay (IDs) and underlay (LOCators). Both spaces have their own separate addresses. Packets use overlay addresses within the sites and are mapped to an underlay address when transmitted across different overlay sites. This separation brings several advantages in terms of new features an scalability of the network (see \cite{ietf-lisp-introduction-13} for further details on this).

Under this principle border routers, that is routers that connect the overlay with the underlay, need to map overlay to underlay address. This mapping is done using the control-plane and is typically cached locally in the border router so that subsequent packets addressed to the same overlay prefix are forwarded directly (fast-path). Several approaches are possible when building Locator/ID split architectures. In this paper we focus on the Locator/ID Separation Protocol  \cite{rfc6830}\cite{ietf-lisp-rfc6830bis-11}\cite{rfc6833}\cite{ietf-lisp-rfc6833bis-09}\cite{7945850}\cite{7158286}. 
In LISP the mapping from the overlay namespaces can be done using two mechanisms. In the first one the mapping is \emph{pulled} from a control-plane infrastructure (referred as Mapping System) using two control-plane messages Map-Request/Map-Reply \cite{rfc6833}. In the second one the mappings use the Publish/Subscribe paradigm \cite{rodrigueznatal-lisp-pubsub-02}, following this architecture xTR subscribe to mappings that are pushed directly from the Mapping System when they are updated. In both cases the mapping is then cached locally in the router (referred as xTRs in LISP terminology).

When pulling the mapping using the control-plane channel attackers can perform a Denial-of-Service or overflow attack against the control-plane channel and/or perform a scanning attack against the local cache. In this paper we discuss these three attacks as well as solutions to address them. The paper uses the LISP terminology but the solutions are applicable to similar ID/LOC split protocols. 
\section{Attacks on the Control-Plane channel and cache}

\subsection{Scenario}

A network is served by a LISP xTR (a LISP-capable router). The LISP site contains both attackers and legitimate users, both are sending data packets towards the xTR. A lookup is performed for each packet against a map cache. The map-cache is assumed to use the Least-Recently Used (LRU) cache replacement policy. 

A hit onto the cache means that the packet is encapsulated and forwarded. For each missed packet the xTR generates a Map-Request, each Map-Request requires storing a nonce (a 64bits nonce in LISP) that is consumed when the corresponding Map-Reply is received. Once the Map-Reply is received a new entry is installed onto the map-cache, this entry is used for subsequent packets addressed towards the same prefix. 

In addition the xTR is equipped with a per-destination EID rate-limiter (e.g., a buffer) for control-plane messages (i.e., Map-Requests). Once the limit is reached subsequent control-plane messages are dropped. The destination EID rate-limiter is assumed to reset after a certain period of time. 

\subsection{Denial-of-Service to the control-plane channel}
This attacks works as follows:
\begin{enumerate}
	\item A (set of) attackers(s) generate(s) packets towards destinations that are not stored in the map-cache. This can be done by targeting non-popular destinations, unallocated IPv6 address space or prefixes that are known not to be part of the overlay address space (negative entries). As an example attackers can use network scanning software for this purpose.
	\item Packets whose destination address is not found in the map-cache generate a control-plane message (Map-Request). Each control-plane message increases the rate-limiter.
	\item Once the rate limit is reached victims cannot communicate with new destinations since subsequent Map-Requests are dropped. In some other cases the attackers may consume more than its fair-share of control-plane messages disrupting the victim's communications.
\end{enumerate}

\subsection{Overflowing the control-plane channel}
This attack is a variant of the previous one and can be mounted using the following steps:
\begin{enumerate}
	\item A (set of) attacker(s) generate(s) packets that are not stored in the map-cache. This can be done by targeting non-popular destinations, unallocated IPv6 address space or prefixes that are known not to be part of the overlay address space (negative entries). As an example attackers can use network scanning software for this purpose.
	\item Packets whose destination address is not found in the map-cache generate a control-plane message (Map-Request). Each outgoing control-plane message requires storing temporary state in the router, in the case of LISP this is a 64-bit random nonce for each control-plane message. This state is consumed once the corresponding Map-Reply message is received.
	\item By generating packets that trigger control-plane messages fast enough the attacker can overflow the memory structure used to temporarily store the nonces.
\end{enumerate}

\subsection{Scanning Attack on the Control-Plane cache}
In this case an attacker can use the well-known scanning attack (or cache pollution) \cite{CONTI20133178}  \cite{4110278} to generate cache evictions thus, disrupting victim’s ongoing flows. The attack works as follows:
\begin{enumerate}
	\item One or multiple attackers send packets over a large period of time (e.g., hours), to destinations having a high probability of not being found in the cache. For instance, this can be achieved by having a list of overlay prefixes (EIDs) and sending packets with destinations enumerating all prefixes in the list in a random order and at a certain packet rate. Once all destinations are exhausted the enumeration would start over. 
	\item Such packets generate control-plane messages that allocate new entries in the cache. If the intensity of the attack is high enough and/or the cache is not large enough the cache will not be able to allocate the new entries. As a result, existing entries will be evicted according to the cache replacement policy, for instance the least recently used. This will result in the disruption of ongoing flows that were using the evicted entries. Such flows will have to pull again the mappings causing packet drops and/or re-routes. 
\end{enumerate}

\section{Solutions to the Attack}

\subsection{Securing the Control-Plane Channel}

In this section we discuss solutions to the attacks described in the previous section. It is important to note that such attacks are well-known by the community since they apply to many fields of IT infrastructures (e.g., processor design \cite{Aad:2004:DSR:1023720.1023741}\cite{4469900}\cite{7081075}). As a consequence different mechanism have been proposed to address such attacks, in this paper we only discuss a subset of the potential solutions.

\subsubsection{Base-line Solution}

First we discuss a base-line solution for this attack: The xTR is equipped with a \emph{per-source} rate-limiter. This rate-limiter counts how many misses each source node (inside the LISP site) has triggered. Once the per-source limit is reached subsequent control-plane messages triggered by this source are dropped. The rate limiter is zeroed after a certain period of time. It is worth noting that this solution assumes that spoofing source addresses is not possible inside the LISP site. 

In order to design such solution the xTR is equipped with a memory structure that counts misses per-source. The size of the structure scales linearly $O(n)$ with the number of nodes within the site (n). This is a very well-known problem in networking and as such several well-established data-structures can be used to build the per-source limiter (e.g., a table or a hash-table\cite{5076211}\cite{Eatherton:2004:TBH:997150.997160}\cite{830044}\cite{912716}) that operate at line-speed for certain values of $n$.

However and if memory size is a concern several more sophisticated data-structures are available (e.g., bloom filters \cite{Song:2005:FHT:1080091.1080114}, count-sketch \cite{Cormode:2008:FFI:1454159.1454225}), such structures while being able to operate at line rate scale (in terms of memory size) \emph{sub-linearly} with the amount of the elements that need to be counted (e.g., nodes inside the LISP site). In what follows we describe one of such structures: the Count-Min Sketch.

\subsubsection{Count-Min Sketch}
The Count-Min Sketch (CMS) \cite{CORMODE200558}\cite{Cormode2012} data structure provides an efficient way to estimate the frequencies of given events. It uses compression methods to scale sub-linearly in size with the amount of elements to count and approximation techniques to estimate its frequencies. When counting the frequencies of an event, CMS applies for each row the corresponding hash function and increments the counter. To estimate the frequency of an event, the minimum of all the counters corresponding to that event is selected. 

\begin{figure}
\includegraphics[height=1.3in, width=3.3in]{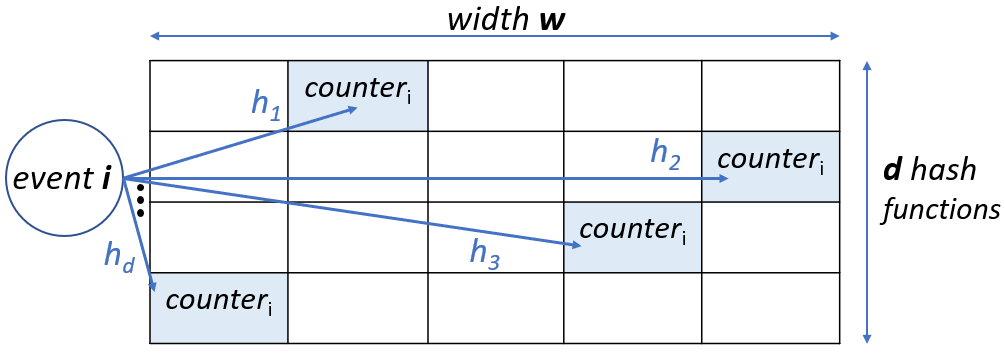}
\caption{Count-Min Sketch mapping to different positions for each row}
\end{figure}

The width of the matrix is assumed to be much smaller than the number of events. The more skewed is the events distribution, the more accurately CMS will approximate the values. All rows are pairwise-independent hash functions that will map an event into a specific position from its corresponding row. The error that a counter value can have is the collision ratio of two or more events falling in the same position. In other words, the sketch will never underestimate the true value of a counter, but it may overestimate it. To decrease the probability of having such error, multiple rows with new independent hash functions are added to the matrix. The following formulas are used to determine the width \textit{w} and the depth \textit{d} of the table, where $\varepsilon$ is the error and $\delta$ is the certainty:
\DeclarePairedDelimiter{\ceil}{\lceil}{\rceil}
\begin{equation}
w = \ceil[\big]{\frac{2}{\varepsilon}}
\end{equation}
\begin{equation}
d = \ceil[\big]{\frac{log{(1-\delta)}}{log{(1/2)}}}
\end{equation}

It is worth noting that CMS is a very popular technique that is being used in production infrastructures, for instance AT\&T as well as Google have deployed it for heavy-hitter detection \cite{10.1007/3-540-45465-9_59}\cite{Cormode:2004:HUS:1007568.1007575} . 

\subsubsection{Preventing Denial-of-Service and overflowing attacks to the control-plane channel}

This solution addresses the attacks defined in sections 2.2 and 2.3 by implementing a per-source rate-limiter taking advantage of the efficiency of the CMS data-structure. An overview of the solution is described in the following figure.
\begin{figure}
\includegraphics[height=1.65in, width=3.4in]{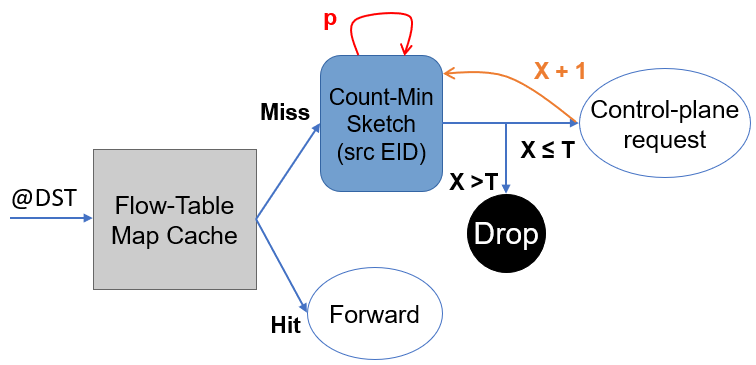}
\caption{Fair share of flow-table misses}
\end{figure}
For each data-packet that misses the map-cache a counter is incremented in the CMS structure using the source address of the node as key (source EID). If counter for that source address exceeds a certain threshold \textit{T} no control-plane message is generated, otherwise a control-plane message is generated. Finally, an aging policy is used to reset all the counters from the data structure after a specific amount of time \textit{p}.

This strategy protects from the previously defined attacks (sections 2.2 and 2.3) since the CMS is used to efficiently detect attackers (heavy-hitters), that is nodes that generate more than \textit{T} control-plane messages each \textit{p} seconds. Attackers are rate-limited to $T/p$ control-plane messages per second.


In order to provide relevant estimates about the size and accuracy of the CMS structure in real scenarios we have performed the following experiments. We use a CMS reference implementation \cite{github}  with a fixed size for each experiment. We consider three network sizes (nodes within the LISP site): 50k, 100k and 500k with two different amount of attackers, 1\% and 10\% are assumed to be attackers. 

Attackers generate (randomly) between 2 and 3 orders of magnitude more control-plane messages than legitimate users. Specifically, attackers generate a uniform random number in the range of 1k-10k, legitimate users a range in 1-10 and the threshold \textit{T} is set to 1k. The size of the CMS structure starts with \textit{w} = 1000 and \textit{d} = 1. The width is being incremented in 1000 in every iteration, and the depth by one unit each 2 iterations. Each cell of the CMS structure consists in a 2-byte counter.

In our experiments we do not consider the reset timer since it is irrelevant with respect to the performance of the CMS structure. Please note that the required CMS size to count the attackers and/or the accuracy does not depend on the specific ratio in the intensity of the control-plane messages generated by attackers and legitimate users, but rather on the ratio of attackers with respect with the legitimate users.

Figure 3 shows the result of the experiments plotting the false positives for different CMS memory sizes and three network scenarios (50k, 100k and 500k). A false positive is when a legitimate user is incorrectly identified as an attacker. As the figure shows the amount of false positive decreases as the size of the CMS increases, this is because in larger structures legitimate users do not collide with attackers. In all three network scenarios and for a certain CMS size the false positive drop to zero, as an example at 56KB for 50k nodes. As a reference a traditional linear structure (e.g., a table) would require at least 100KB of memory. If an error lower than 0.1\% is acceptable, then size of the CMS structure drops to 30KB for 50k nodes. We also plot in figure 4 an scenario with 10\% of attackers, in this case the CMS structure requires a size of 288KB for 50k nodes with zero error and 168KB for errors below 0.1\%. It is worth noting that the CMS structure does not underestimate the values and as such false negative are negligible. 

\begin{figure}
\includegraphics[width=\columnwidth]{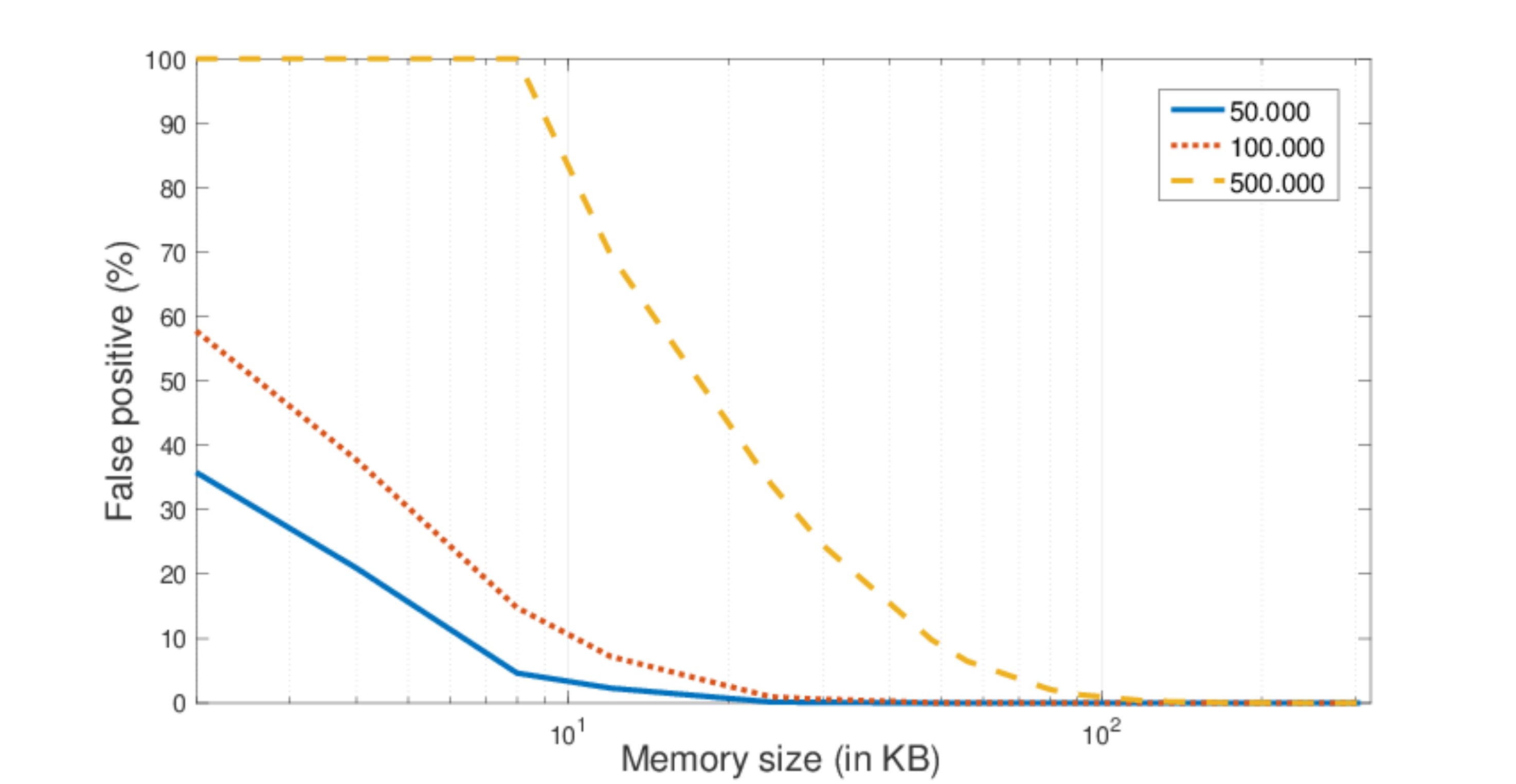}
\caption{False positive detection of attacks as a function of the CMS size (KB) when 1\% of nodes are attackers}
\end{figure}
\begin{figure}
\includegraphics[width=\columnwidth]{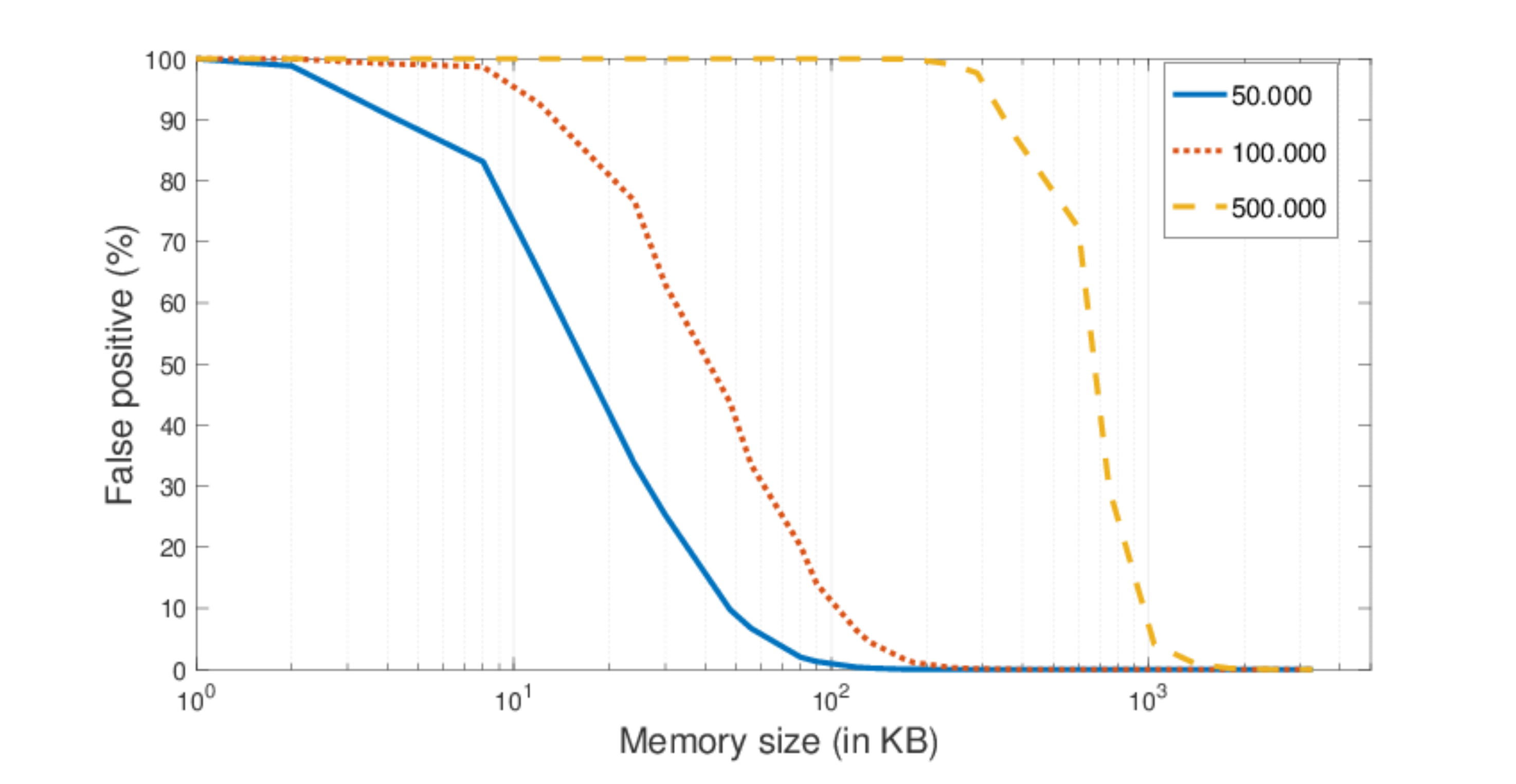}
\caption{False positive detection of attacks as a function of the CMS size (KB) when 10\% of nodes are attackers}
\end{figure}

Finally it is worth noting that similar structures considering commensurate sizes have been implemented in switches operating at line-rate. We refer the interested reader to  \cite{Sivaraman:2017:HDE:3050220.3063772} for further information on heavy hitter detection.

\subsection{Securing the Control-Plane Cache}

Cache scanning attacks are well-known in the IT field and as a result a wide plethora of solutions have been proposed for this. In particular cache replacement policies have been designed to specifically mitigate such attacks. Notable examples are the Least Frequently Used (LFU)-Aging \cite{Arlitt:2000:ECM:346000.346003} \cite{DBLP:journals/corr/abs-1001-4135}  \cite{Robinson:1990:DCM:98460.98523},  ARC \cite{Megiddo:2003:ASL:1090694.1090708},  Least Recently Frequently Used (LRFU)\cite{Lee:1999:ESP:301464.301487}\cite{Lee:2001:LSP:626527.627193} and the Least Frequent Recently Used (LFRU)\cite{7857720}. 

As an example a LFU-A \cite{DBLP:journals/corr/abs-1001-4135}\cite{Robinson:1990:DCM:98460.98523} replacement policy works by evicting the cache entry with less hits (less packets addressed towards this destination) when the cache is full. In addition and after a certain amount of time the reference counts for  popular entries is decremented (or zeroed) to make them candidates for replacement.

Under such policy attackers must generate as much traffic as the traffic addressed to popular destinations in order to disrupt them. We refer the interested reader to \cite{6975256} for further details about this topic.

\section{Discussion}

The main design rationale behind the proposed solution is to detect and push-back attackers by rate-limiting them in aggregating points. This is a common practice in network operation that seeks mitigating security threats as topologically close as possible to the attackers. In what follows we discuss the main assumptions of this solution: 

\textbf{Spoofing:} The security mechanism proposed in this paper assumes that the network serviced by the xTR has deployed spoofing prevention mechanisms. This is as a reasonable assumption in managed IT infrastructures such as 5G or Data-Center scenarios. If the network does not deploy anti-spoofing mechanisms then a wide-range of attacks are possible, in some scenarios such attacks attacks are more disruptive than the one described in this paper.

\textbf{What is the source in the per-source rate-limiter? } The proposed solution is assumed to count source IP addresses, however the solution can be trivially adapted to count any other field of the packets such as MAC address or information provided by the network infrastructure (e.g, Ethernet ports or wireless access point information). This enables the solution to count misses with different granularities.

\textbf{On the ratio of attackers:} The performance of the CMS depends on the ratio of attackers with respect to the size of the nodes serviced by the xTR. Typically it is assumed that the attackers represent a small portion of the overall population, this results in skewed statistical distributions that enable efficient use of the resources (e.g, CMS, caching, etc). If this assumption does not hold then the resources have to be provisioned for the worst-case (e.g, 100\% of the nodes are attackers). This means that some elements of the data-plane must store the full network state resulting in inefficient use of the resources as well as increased CAPEX. 

In short, if these assumptions do not hold alternative LISP deployment models can be used where some LISP elements store all the network state \cite{draft-rodrigueznatal-ila-lisp-00}.

\section{Summary}
In this short paper we have described two attacks to the control-plane channel and cache of pull-based ID/LOC split protocols. The first attacks aims to DoS or overflow the control-plane channel, for this attack we have described a solution that implements a per-source rate-limiter that takes advantage of the CMS structure, this structure operates at line-rate and scales sub-linearly with the amount of nodes in terms of size. The second attack is a scanning attack against the control-plane cache, this is a well-known topic and we have pointed the interested reader to a set of already existing solutions. Finally in the paper we have borrowed the LISP (RFC6830) terminology but both the attacks and the solutions can be applied to any pull-based ID/LOC split protocol.

\begin{acknowledgments}
This work has been partially supported by the Spanish Ministry of Economy and Competitiveness under contract TEC2017-90034-C2-1-R (ALLIANCE project) that receives funding from FEDER and by the Catalan Institution for Research and Advanced Studies (ICREA).
\end{acknowledgments}

\bibliography{references}
\end{document}